\documentclass[showpacs,preprintnumbers,amsmath,amssymb,floatfix,eqsecnum]{revtex4}

\usepackage{color}   
\usepackage{graphicx}
\usepackage{dcolumn}
\usepackage{bm}

\begin{document}

\def\bbox#1{\hbox{\boldmath${#1}$}}
\def\blambda{{\hbox{\boldmath $\lambda$}}}
\def\eeta{{\hbox{\boldmath $\eta$}}}
\def\bxi{{\hbox{\boldmath $\xi$}}}
\def\bzeta{{\hbox{\boldmath $\zeta$}}}

\title{ Landau Hydrodynamics Reexamined 
    }

\author{Cheuk-Yin Wong\footnote{wongc@ornl.gov}}

\affiliation{Physics Division, Oak Ridge National Laboratory, Oak
Ridge, TN  37831}

\date{\today}

\begin{abstract}

We review the formulation of Landau hydrodynamics and find that the
rapidity distribution of produced particles in the center-of-mass
system should be more appropriately modified as $dN/dy \propto \exp \{
\sqrt{y_b^2-y^2}\}$, where $y_b=\ln \{\sqrt{s_{NN}}/m_p\}$ is the beam
nucleon rapidity, instead of Landau's original distribution,
$dN/dy({\rm Landau}) \propto \exp \{ \sqrt{L^2-y^2}\}$, where $L=\ln
\{\sqrt{s_{NN}}/2m_p\}$.  The modified distribution agrees better with
experimental $dN/dy$ data than the original Landau distribution and
can be represented well by the Gaussian distribution, $dN/dy({\rm
Gaussian}) \propto \exp \{ -y^2/2L\}$.  Past successes of the Gaussian
distribution in explaining experimental rapidity data can be
understood, not because it is an approximation of the original Landau
distribution, but because it is in fact a close representation of the
modified distribution.  Predictions for pp and AA collisions at LHC
energies in Landau hydrodynamics are presented.

\end{abstract}


\maketitle


\large
 \section {\bf 
Introduction}
\vspace{0.3cm}

Recent experimental data in high-energy heavy-ion collisions
\cite{Mur04,Ste05,Ste07} reveal that the rapidity distributions of
produced particles do not exhibit the plateau structure of Hwa-Bjorken
hydrodynamics \cite{Hwa74,Bjo83}.  On the contrary, the Landau
hydrodynamical model \cite{Lan53,Bel56} yields results that agree with
experiment \cite{Mur04,Ste05,Ste07}.  Landau hydrodynamics provides a
plausible description for the evolution of the dense hot matter
produced in high-energy heavy-ion collisions.  Its dynamics during the
first stage of the one-dimensional longitudinal expansion can be
solved exactly and the one-dimensional longitudinal expansion problem
admits simple approximate solutions
\cite{Lan53,Bel56,Kha54,Bel56a,Ama57,Car73,Coo74,Coo75,Cha74,Sri92,Moh03,Ham05,Pra07,Bia07,Cso08,Beu08,Osa08}.
The subsequent three-dimensional motion can be solved approximately to
give rise to predictions that come close to experimental data
\cite{Lan53,Bel56,Mur04,Ste05,Ste07}.  A critical re-examination of
Landau hydrodynamics will make it a useful tool for the description of
the evolution of the produced dense matter.

Quantitative analyses of Landau hydrodynamics in
Refs. \cite{Mur04,Ste05,Ste07,Car73} use a Gaussian form of the Landau
rapidity distribution \cite{Lan53,Bel56}
\begin{eqnarray}
\label{gau}
dN/dy({\rm Gaussian}) \propto \exp\{-y^2/2L\},
\end{eqnarray}
where $L$ is the logarithm of the Lorentz contraction factor
$\gamma=\sqrt{s_{NN}}/2m_p$,
\begin{eqnarray}
L=\ln \gamma= \ln (\sqrt{s_{NN}}/2m_p),
\end{eqnarray}
 $\sqrt{s_{_{NN}}}/2$ is the center-of-mass energy per nucleon, and
$m_p$ is the proton mass.  This Gaussian rapidity distribution gives
theoretical rapidity widths that agree with experimental widths for
many different particles in central AuAu collisions, to within 5 to
10\%, from AGS energies to RHIC energies \cite{Mur04,Ste05,Ste07}.
The Landau hydrodynamical model also gives the correct energy
dependence of the observed total charged multiplicity and the limiting
fragmentation property at forward rapidities \cite{Ste05,Ste07}.  A
similar analysis in terms of the pseudorapidity variable $\eta$ at
zero pseudorapidity has been carried out in \cite{Sar06}.

The successes of these analyses indicate that Landau hydrodynamics can
be a reasonable description.  However, they also raise many unanswered
questions.  Firstly, the original Landau result stipulates the
rapidity distribution to be \cite{Lan53,Bel56}
\begin{eqnarray}
\label{Lan}
dN/d\lambda({\rm Landau}) \propto \exp \{ \sqrt{L^2-\lambda^2}\},
\end{eqnarray}
where the symbol $\lambda$ is often taken to be the rapidity variable
$y$ in \cite{Mur04,Ste05,Ste07,Car73}.  In the original work of Landau
and his collaborator in \cite{Lan53,Bel56}, the variable $\lambda$ is
used to represent the polar angle $\theta$ as $e^{-\lambda}=\theta$;
there is the question whether the variable $\lambda$ in the Landau
rapidity distribution (\ref{Lan}) should be taken as the rapidity
variable $y$ \cite{Mur04,Ste05,Ste07,Car73} or the pseudorapdity
variable $\eta$ \cite{Sar06} appropriate to describe the polar angle.
Such a distinction between the rapidity and pseudorapidity variables
is quantitatively important because the shape of the distributions in
these two variables are different near the region of small rapidities
\cite{Won94}.  Secondly, the Gaussian rapidity distribution
(\ref{gau}) used in the analyses of Refs.\ \cite{Mur04,Ste05,Ste07} is
only an approximate representation of the original Landau distribution
(\ref{Lan}) in the region of $|\lambda| \ll L$, but differs from the
original Landau distribution (\ref{Lan}) in other rapidity regions.
They are in fact different distributions.  While the original Landau
distribution can be considered to receive theoretical support in
Landau hydrodynamics as justified in Refs. \cite{Lan53,Bel56}, a firm
theoretical foundation for the Gaussian distribution (\ref{gau}) in
Landau hydrodynamics is still lacking.  Finally, if one does not use
the approximate representation of the Gaussian distribution
(\ref{gau}) but keeps the original Landau distribution (\ref{Lan}),
then there is the quantitative question \cite{Cha74} whether this
original Landau distribution will give results that agree with
experimental data.

In view of the above unanswered questions, our task in reviewing the
Landau hydrodynamical model will need to ensure that we are dealing
with the rapidity variable $y$ and not the pseudorapidity variable
$\eta$.  We need to be careful about various numerical factors so as
to obtain a quantitative determination of the parameters in the final
theoretical results.  Finally, we need to ascertain whether the
theoretical results agree with experimental data.  If we succeed in
resolving the unanswered questions, we will pave the way for the
application of Landau hydrodynamics to other problems in high-energy
heavy-ion collisions.

\vspace*{0.5cm} \section {\bf Total Number of Produced Charged
Particles }
\vspace{0.3cm}

Landau hydrodynamics involves two different aspects: the global
particle multiplicity and the differential rapidity distribution.
Landau assumed that the hydrodynamical motion of the fluid after the
initial collision process is adiabatic.  He argued that the only thing
that can destroy adiabaticity would be the shock waves which however
occur at the initial compressional stage of the collision process
\cite{Won74}.  Landau therefore assumed that during the longitudinal
and transverse expansion phase under consideration, the entropy
content of the the individual region remains unchanged.  The total
entropy of the system is therefore unchanged and can be evaluated at
the initial stage of the overlapped and compressed system.

From the consideration of the thermodynamical properties of many
elementary systems, Landau found that the ratio of the entropy density
to the number density for a thermally equilibrated system is nearly a
constant within the temperature regions of interest.  Landau therefore
postulated that the number density is proportional to the entropy
density. Thus, by collecting all fluid elements, the total number of
particles is proportional to the total entropy.  As the total entropy
of the system is unchanged during the hydrodynamical evolution, the
total number of observed particles can be determined from the initial
entropy of the system.

We work in the center-of-mass system and consider the central
collision of two equal nuclei, each of mass number $A$, at a
nucleon-nucleon center-of-mass energy $\sqrt{s_{_{NN}}}$.  Consider
first the case of central AA collisions with $A \gg 1$ such that
nucleons of one nucleus collide with a large numbers of nucleons of
the other nucleus and the whole energy content is used in particle
production.  The total energy content of the system is
\begin{eqnarray}
\label{Etot}
E=\sqrt{s_{_{NN}}}A.
\end{eqnarray}
The initial
compressed system is contained in a volume that is Lorentz contracted
to become
\begin{eqnarray}
V=\frac{4\pi}{3} (r_0 A^{1/3}) ^3/\gamma ,
\end{eqnarray}
where $r_0=1.2$ fm.  The energy density of the system is therefore
\begin{eqnarray}
\epsilon = E/V= \gamma \sqrt{s_{_{NN}}}/(4\pi r_0^3/3).
\end{eqnarray}
For a system in local thermal equilibrium, the entropy density
$\sigma$ is related to the energy density by
\begin{eqnarray}
\sigma = {\rm constant~}\epsilon^{3/4}.
\end{eqnarray}
The total entropy content of the system is therefore
\begin{eqnarray}
S=\sigma V ={\rm ~constant~} s_{_{NN}}^{1/4} A.  
\end{eqnarray}
With Landau's assumption relating entropy and particle number, 
$N\propto S$, the total number of particles produced is
\begin{eqnarray}
N \propto s_{_{NN}}^{1/4} A,  
\end{eqnarray}
and the total number of produced charged particles per participant
pair is
\begin{eqnarray}
\label{ncha}
N_{\rm ch}/A = 
N_{\rm ch}/(N_{\rm part}/2) = K (\sqrt{s_{_{NN}}}/{\rm GeV})^{1/2}, 
\end{eqnarray}
where $K$ can be determined phenomenologically by comparison with
experimental data.

\begin{figure} [h]
\includegraphics[angle=0,scale=0.50]{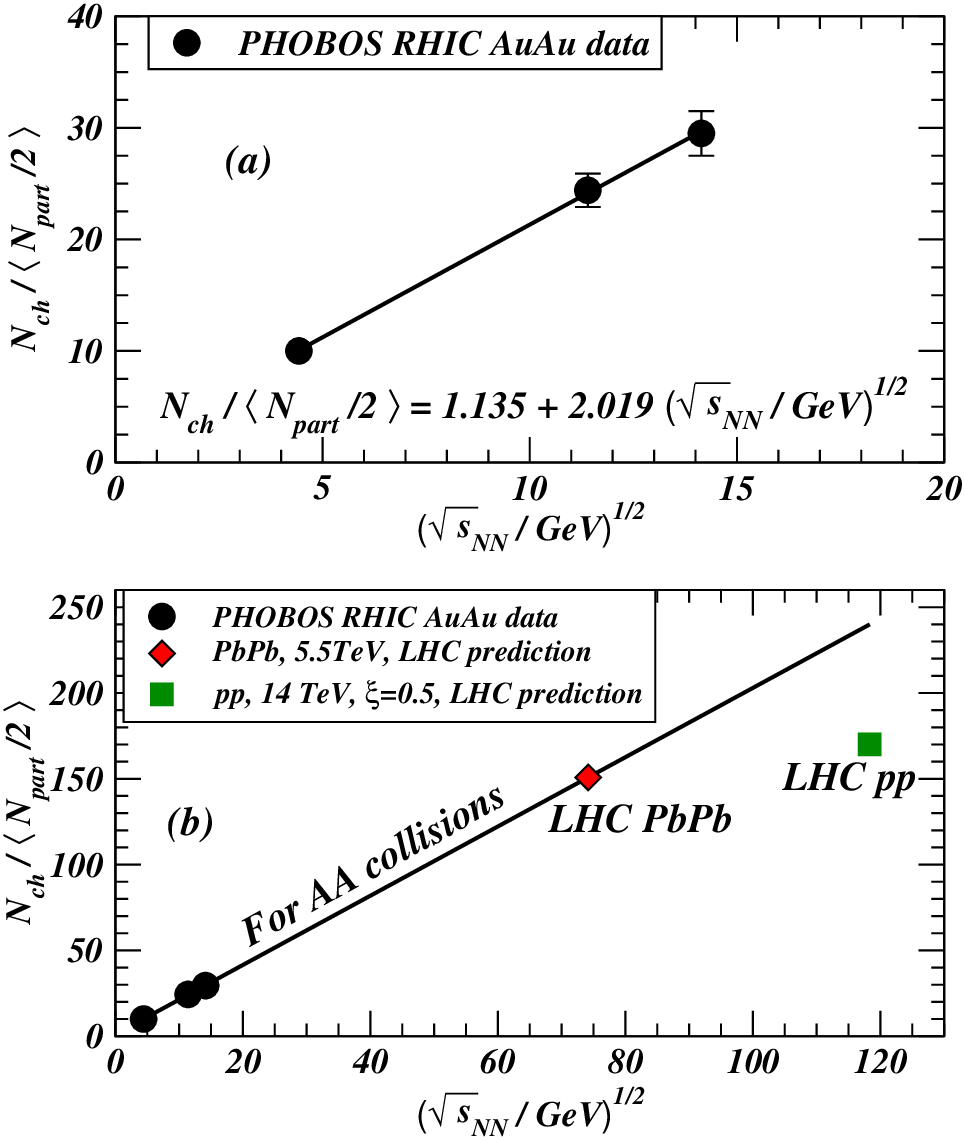}
\vspace*{0.0cm} 
\caption{(Color online) Total number of produced charged particles per
pair of participants, $N_{\rm ch}/(N_{\rm part}/2)$, as a function of
$(\sqrt{s_{_{NN}}}/{\rm GeV})^{1/2}$.  (a) PHOBOS data $N_{\rm
ch}/(N_{\rm part}/2)$ data for central AuAu collisions at different
$(\sqrt{s_{_{NN}}}/{\rm GeV})^{1/2}$ and the Landau hydrodynamical
model fit, and (b) the extrapolation of the charged multiplicity in
Landau hydrodynamical model to pp and PbPb collisions at LHC
energies.}
\end{figure}

In Fig.\ 1(a), we show the PHOBOS data of $N_{\rm ch}/(N_{\rm
  part}/2)$ as a function of $(\sqrt{s_{{NN}}}/{\rm GeV})^{1/2}$ for
  central AuAu collisions in RHIC \cite{Ste05,Ste07}.  The RHIC
  AuAu data can be parametrized as
\begin{eqnarray}
\label{system}
N_{\rm ch}/(N_{\rm part}/2) = 1.135 + 2.019 
(\sqrt{s_{_{NN}}}/{\rm GeV})^{1/2},
\end{eqnarray}
where the constant 1.135 arise from the leading baryons.  The
constant $K$ as determined from the data is $K= 2.019$ which agrees
with the earlier estimate of $K=2$ \cite{Lan53,Bel56}.  

Consider next pp and p\= p collisions in which not all the energy of
$\sqrt{s_{_{NN}}}$ is used in particle production, as the leading
particles carry a substantial fraction of the initial energy.  If we
denote the particle production energy fraction in pp and p\= p
collisions by $\xi$, then Eq.\ (\ref{ncha}) is modified to be
\begin{eqnarray}
N_{\rm ch}= 
K (\xi \sqrt{ s_{_{NN}}}/{\rm GeV})^{1/2}. 
\end{eqnarray}
Comparison of the charged particle multiplicity in pp and p\= p
collisions indicates that the particle production energy fraction
$\xi$ for pp and p\= p collisions is approximately 0.5
\cite{Bas81,Won94,Ste05,Ste07}.  In contrast, the case of RHIC AA data
in high-energy heavy-ion collisions corresponds to full nuclear
stopping with $\xi=1$ \cite{Ste05,Ste07}.

In Fig.\ 1(b), we show the predictions for the charge particle
multiplicity per pair of participants for collisions at LHC energies.
For pp collisions at 14 TeV with a particle production energy fraction
$\xi=0.5$, $N_{\rm ch}$ is predicted to be 170.  For central PbPb
collisions at $\sqrt{s_{_{NN}}}= 5.5$ TeV with full nuclear stopping
$(\xi=1)$, $N_{\rm ch}/(N_{\rm part}/2)$ is predicted to be 151.

\vspace*{0.5cm} \section {\bf Longitudinal Hydrodynamical Expansion }
\vspace{0.3cm}

We proceed to examine the dynamics of the longitudinal and transverse
expansions in the collision of two equal nuclei of diameter $a$.  The
disk of initial configuration in the center-of-mass system has a
longitudinal thickness $\Delta$ given by
\begin{eqnarray}
\label{eq11}
\Delta = a /\gamma,
\end{eqnarray}
as depicted in Fig.\ 2 with major diameters $a_x$ and $a_y$ and the 
reaction plane lying on the $x$-$z$ plane.  Depending on the impact
parameter, the dimensions of the disk obey $a_x \le a_y \le a$.  For a
central collision, $a_x = a_y = a$.

\begin{figure} [h]
\includegraphics[angle=0,scale=0.50]{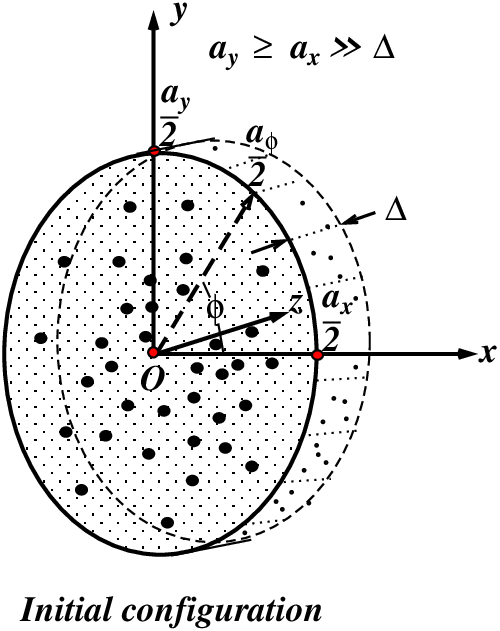}
\vspace*{0.0cm} 
\caption{Initial configuration in the collision of two heavy equal
  nuclei in the center-of-mass system.  The region of nuclear overlap
  consists of a thin disk of thickness $\Delta$ along the longitudinal
  $z$-axis. The reaction plane is designated to lie on the $x$-$z$
  plane, and the transverse radii are $a_x/2$ and $a_y/2$.}
\end{figure}

Among the coordinates $(t,z,x,y)\equiv (x^0,x^1,x^2,x^3)$ used to
describe the fluid, Landau suggested a method to split the problem
into two stages.  The first stage consists of independent expansions
along the longitudinal and the transverse directions.  For the
longitudinal expansion, the equation of hydrodynamics is
\begin{eqnarray}
\label{T00}
\frac{\partial T^{00}}{\partial t} + \frac{\partial T^{01}}{\partial z} =0,\\
\label{T01}
\frac{\partial T^{01}}{\partial t} + \frac{\partial T^{11}}{\partial z} =0,
\end{eqnarray}
where
\begin{eqnarray}
T^{\mu \nu}=(\epsilon + p)u^\mu u^\nu - p g^{\mu \nu}.
\end{eqnarray}
We shall assume for simplicity the relativistic equation of state
\begin{eqnarray}
\label{eqst}
p=\epsilon/3.
\end{eqnarray}
To ensure that we deal with rapidities,
we represent the velocity fields $(u^0, u^1)$ by the
flow rapidity $y$
\begin{subequations}
\begin{eqnarray}
\label{u0}
u^0&=& \cosh y, \\
\label{u1}
u^1&=& \sinh y.
\end{eqnarray}
\end{subequations}
We introduce the light-cone coordinates $t_+$ and $t_-$
\begin{subequations}
\begin{eqnarray}
t_+&=& t+z, \\
t_-&=& t-z, 
\end{eqnarray}
\end{subequations}
with their logarithmic representations  $(y_+,y_-)$ defined by
\begin{eqnarray}
\label{ypm}
y_{\pm} = \ln \{ t_{\pm} /\Delta \} = \ln \{ (t {\pm} z) /\Delta \}.
\end{eqnarray}
The hydrodynamical equations (\ref{T00}) and
(\ref{T01}) become
\begin{subequations}
\begin{eqnarray}
\label{eq1da}
\frac{\partial \epsilon}{\partial t_+} 
+ 2 \frac{\partial (\epsilon e^{-2y}) }{\partial t_-}&=&0,\\
\label{eq1db}
2 \frac{\partial (\epsilon e^{2y}) }{\partial t_+} 
+  \frac{\partial \epsilon  }{\partial t_-}&=&0.
\end{eqnarray}
\end{subequations}

For the first stage of one-dimensional hydrodynamics, the exact
solution for an initially uniform slab has been obtained and discussed
in
\cite{Bel56,Kha54,Bel56a,Ama57,Car73,Coo74,Coo75,Cha74,Sri92,Moh03,Ham05,Pra07,Bia07,Cso08,Beu08}.
There are in addition simple approximate solutions \cite{Lan53,Bel56}.
In view of the matching of the solution to an approximate
three-dimensional motion in the second stage, it suffices to consider
the approximate solutions given by \cite{Bel56}
\begin{subequations}
\begin{eqnarray}
\label{sole}
\epsilon (y_+,y_-) & = &  \epsilon_0\exp\left \{ -\frac{4}{3}(y_+ + y_- -
\sqrt{y_+y_-}) \right \},\\
\label{soly}
y (y_+,y_-) & = &  (y_+- y_-)/2.
\end{eqnarray}
\end{subequations}
The flow rapidity equation of Eq.\ (\ref{soly}) can also be written alternatively as
\begin{eqnarray}
\label{sol2}
e^{ 2 y(y_+,y_-) } = \frac{t_+}{t_-} = \frac{t + z}{t-z}.
\end{eqnarray}
The constant $\epsilon_0$ in Eq.\ (\ref{sole}) is related to the
initial energy density at $(y_{+0}, y_{-0})$ by
\begin{eqnarray}
\epsilon_0=
\epsilon(y_{+0},y_{-0}) e^{\phi_0},
\end{eqnarray}
where $\phi_0$ is 
\begin{eqnarray}
\phi_0= \frac{4}{3} ( y_{+0} +y_{-0}
-\sqrt{y_{+0}y_{-0}}).
\end{eqnarray}
We can easily prove by direct substitution that (\ref{sole}) and
(\ref{soly}) (or (\ref{sol2})) are approximate solutions of the
hydrodynamical equations (\ref{eq1da}) and (\ref{eq1db}).  First,
substituting Eq.\ (\ref{sol2}) into the hydrodynamical equations, we
obtain
\begin{subequations}
\begin{eqnarray}
\frac{\partial \epsilon}{\partial t_+} 
+ 2 \left [ \frac{\partial \epsilon  }{\partial t_-}
+\frac{\epsilon}{t_-} \right ] \frac{t_-}{t_+}
&=&0,\\
2  \left [\frac{\partial \epsilon  }{\partial t_+} 
+\frac{\epsilon}{t_+} \right ]
 \frac{t_+}{t_-}
+  \frac{\partial \epsilon  }{\partial t_-}&=&0.
\end{eqnarray}
\end{subequations}
We write out $t_-/t_+$ in the second equation and substitute it into
the first equation, and we get
\begin{eqnarray}
\frac{\partial \epsilon}{\partial t_+} 
\frac{\partial \epsilon}{\partial t_-} 
- 4 \left [ \frac{\partial \epsilon  }{\partial t_-}
+\frac{\epsilon}{t_-} \right ] 
  \left [\frac{\partial \epsilon  }{\partial t_+} 
+\frac{\epsilon}{t_+} \right ]
=0.
\end{eqnarray}
We multiply this expression by $t_+ t_-$ and change into the
logarithm variables $y_+$ and $y_-$, then the above equation becomes
\begin{eqnarray}
\frac{\partial \epsilon}{\partial y_+} 
\frac{\partial \epsilon}{\partial y_-} 
- 4 \left [ \frac{\partial \epsilon  }{\partial y_-}
+ \epsilon \right ] 
  \left [\frac{\partial \epsilon  }{\partial y_+} +\epsilon \right ]
=0
\end{eqnarray}
If we now substitute Eq.\ (\ref{sole}) for $\epsilon$ into the
lefthand side of the above equation, we find that the lefthand side
gives zero, indicating that Eqs.\ (\ref{sole}) and (\ref{soly}) are
indeed approximate  solutions of the hydrodynamical equation.

The simple approximate solutions of (\ref{sole}) and (\ref{soly}) have
limitations.  They cannot describe the boundary layers for which
$|t\pm z| <\Delta$ and $y_\pm $ becomes negative.  In highly
relativistic collisions, the tail regions excluded from the
approximate solution are not significant in a general description of
the fluid.  The solutions in (\ref{sole}) and (\ref{soly}) provide
only limited choice on the initial conditions, within the form as
specified by the simple functions in these equations.  However, a thin
slab of matter with the right dimensions within the Landau model will
likely capture the dominant features of the evolution dynamics.

It is useful to compare Landau hydrodynamics with Hwa-Bjorken
hydrodynamics.  We make the transformation $t= \tau \cosh y,$ and $z=
\tau \sinh y$.  The energy density is then
\begin{eqnarray}
\epsilon (\tau, y)  =  \epsilon_0\exp\left \{ 
-\frac{4}{3}\left [ 2\ln({\tau}/{\Delta})-\sqrt{[\ln (\tau/\Delta)]^2-y^2}
\right ] \right \}.
\end{eqnarray}
In the  region $y\ll \ln (\tau/\Delta)$, we have
\begin{eqnarray}
\epsilon (\tau, y)  \sim \epsilon_0\exp\left \{
-\frac{4}{3}\ln({\tau}/{\Delta}) \right \} \propto \frac{1}{\tau^{4/3}},
\end{eqnarray}
which is the Hwa-Bjorken hydrodynamics results.  Therefore, in the
region of small rapidities with $|y| \ll\ln (\tau/\Delta)$, Landau
hydrodynamics and Hwa-Bjorken hydrodynamics coincide.  In general,
because Landau hydrodynamics covers a wider range of rapidities which
may not be small, it is a more realistic description for the evolution
of the hydrodynamical system.

\large
\vspace*{0.5cm}
\section {\bf 
Transverse Expansion }
\vspace{0.3cm} 

The initial configuration is much thinner in the longitudinal
direction than in the transverse directions. Therefore, in the first
stage of the evolution during the fast one-dimensional longitudinal
expansion, there is a simultaneous but slower transverse expansion.
The difference in the expansion speeds allows Landau to treat the
longitudinal and transverse dynamics as independent expansions.  The
rate of transverse expansion can then be obtained to provide an
approximate description of the dynamics of the system.

We shall consider first the case of a central collision, for which
$a_y=a_x=a$.  The case of non-central collisions will be discussed in
Section IX.  The transverse expansion is governed by the Euler equation
along one of the transverse directions, which can be taken to be along
the $x$ direction,
\begin{eqnarray}
\label{T02}
\frac{\partial T^{02}}{\partial t} 
+ \frac{\partial T^{22}}{\partial x} =0,
\end{eqnarray}
where 
\begin{eqnarray}
T^{02}=(\epsilon+p) u^0 u^2=\frac{4}{3} \epsilon u^0 u^0 v_x,
\end{eqnarray}
and we have used the relation $u^2=u^0 v_x$.  The energy-momentum
tensor $T^{22}$ is
\begin{eqnarray}
T^{22}=(\epsilon+p) u^2 u^2 -p g^{22}=\frac{4}{3} \epsilon u^0 u^0
v_x v_x +p.
\end{eqnarray}
As the transverse expansion is relatively slow, we can neglect the
first term on the righthand side of the above expression and keep only
the pressure term $p$.

In Landau's method of splitting the equations, one makes the
approximation that during the first stage the quantities $\epsilon$
and $y$ as a function of $t$ and $z$ have been independently
determined in the one-dimensional longitudinal motion.  
Equation\ (\ref{T02}) can therefore be approximated
as
\begin{eqnarray}
\label{trans}
\frac{4}{3} \epsilon u^0 u^0 \frac{\partial v_x}{\partial t}
=-\frac{\partial p}{\partial x}.
\end{eqnarray}
The transverse displacement $ x(t)$ (relative to zero displacement) as
a function of time $t$ is related to the acceleration $\partial
v_z/\partial t$ by
\begin{eqnarray}
 x (t) = 
 \frac{1}{2}\left ( \frac{\partial v_x}{\partial t} \right ) t^2. 
\end{eqnarray}
The pressure is $p=\epsilon/3$ at the center of the transverse region and
is zero at the radial surface $a/2$.  Therefore the equation for the
displacement is given from Eq.\ (\ref{trans}) by
\begin{eqnarray}
\label{displ}
\frac{4}{3} \epsilon u^0 u^0 \frac{ 2 x(t) }{ t^2}
= \frac{ \epsilon}{ 3a/2}. 
\end{eqnarray}
We note that there is a factor of 4 arising from the ratio of
$4\epsilon /3$ from $(\epsilon+p)$ on the lefthand side relative to
$\epsilon/3$ from the pressure $p$ on the righthand side.  However, in
the original formulation of Landau \cite{Lan53,Bel56}, this factor of
4 is taken to be unity for an order of magnitude estimate of the
transverse displacement.  For our purpose of making quantitative
comparison with experimental data, this factor of 4 cannot be
neglected. 

From Eq.\ (\ref{displ}), the transverse displacement $x(t)$ during
the longitudinal expansion increases
as
\begin{eqnarray}
\label{xt}
 x(t) = \frac{t^2}{4 a u^0 u^0 }=\frac{t^2}{4 a \cosh^2 y} .
\end{eqnarray}

\large
\vspace*{0.5cm}

\section {\bf 
Second Stage of Conic Flight}
\vspace{0.3cm} 

Landau suggested that when the transverse displacement $x(t)$ is equal
to $a$ at $t=t_{\rm FO}$, we need to switch to a new type of solution in the
second stage of fluid dynamics.  With the fluid element beyond the
initial transverse dimension, hydrodynamical forces become so small
that they can be neglected in the hydrodynamical equations at these
locations and the flow rapidity $y$ can be assumed to be frozen for
$t\ge t_{\rm FO}$.  This is equivalent to freezing the opening polar angle
$\theta$ between the fluid trajectory and the collision axis.  The
motion of the fluid element with a fixed polar angle can be described
as a `three-dimensional' conic flight.  In mathematical terms,
Landau's condition for rapidity freeze-out occurs at $t_{\rm FO}(y) $ which
satisfies \cite{Lan53,Bel56}
\begin{eqnarray}
\label{xtt}
 x(t_{\rm FO}) = a.
\end{eqnarray}
As determined from Eqs.\ (\ref{xt}) and (\ref{xtt}), rapidity
freeze-out takes place at
\begin{eqnarray}
\label{tmy}
t_{\rm FO}(y) = 2 a u^0 =2 a \cosh y. 
\end{eqnarray}
The set of the $(t_{\rm FO}(y),y)$ points lie on the curve of
the proper time, $\tau_{\rm FO}=2a$.  Thus, Landau's physical freeze-out
condition, Eq.\ (\ref{xtt}), corresponds to particle rapidities
freezing-out at a fixed proper time,
\begin{eqnarray}
\tau_{\rm FO} = 2 a.
\end{eqnarray}
In a conic flight with an opening polar angle $\theta$ within
an angle element $d\theta$, the energy-momentum
tensor and the entropy flux within the cone element must be conserved
as a function of time. The cross sectional area of such a cone element
is $2\pi x dx$.  So the conservation of energy and entropy conic flow
correspond to
\begin{eqnarray}
\label{de}
dE=\epsilon u^0 u^0 2\pi x dx = {\rm ~constant},
\end{eqnarray}
and
\begin{eqnarray}
dS=\sigma u^0 2\pi x dx = \epsilon^{3/4} u^0 2\pi x dx = {\rm ~constant}.
\end{eqnarray}
Dividing the first equation by
the second equation, we get
\begin{eqnarray}
\epsilon^{1/4} u^0= {\rm ~constant},
\end{eqnarray}
which gives
\begin{eqnarray}
\label{ett}
\epsilon \propto \frac{1}{(u^0)^4}.
\end{eqnarray}
On the other hand, in the conic flight, $x$ and $dx$ are proportional
to $t$. Hence, Eq.\ (\ref{de}) gives
\begin{eqnarray}
\label{uutt}
\epsilon u^0u^0 t^2 = {\rm ~constant}.
\end{eqnarray}
Eqs.\ (\ref{ett}) and (\ref{uutt}) yield the dependence of various
quantities as a function of $t$,
\begin{eqnarray}
\epsilon \propto \frac{1}{t^4},
~~~\sigma \propto \frac{1}{t^3}, {\rm ~~~and~~~~}
u^0 \propto t.
\end{eqnarray}
These equations give the solution of the evolution of the fluid
elements as a function of time in the second stage.  By matching the
solutions at $t=t_{\rm FO}(y)$, the energy density and velocity fields at the
second stage for $t \ge t_{\rm FO}(y)$ is
\begin{subequations}
\begin{eqnarray}
\epsilon (t,y)& =& 
\epsilon (t_{\rm FO},y)\, {t_{\rm FO}^4}/{t^4}\\
u^0(t,y)& = & u^0(t_{\rm FO},y)\, {t}/{t_{\rm FO}}.
\end{eqnarray}
\end{subequations}

\large
\vspace*{0.5cm} 
\section{\bf Rapidity Distributions
in High Energy Heavy-Ion Collisions }
\vspace{0.3cm}

The picture that emerges from Landau hydrodynamics can be summarized
as follows.  For an initial configuration of a thin disk of dense
matter at a high temperature and pressure, the first stage of the
motion is a one-dimensional longitudinal expansion with a simultaneous
transverse expansion.  The transverse expansion lead to a transverse
displacement.  When the magnitude of the transverse displacement
exceeds the initial transverse dimension, forces acting on the fluid
element becomes small and the fluid elements will proceed to the
second stage of conic flight with a frozen rapidity.  As the
transverse displacement depends on rapidity, and the transverse
displacement magnitude decreases with increasing rapidity magnitude,
the moment when the fluid element switches from the first stage to the
second stage depends on the rapidity.  The final rapidity distribution
of particles is therefore given by the rapidity distribution of the
particles at the matching time $t_{\rm FO}(y) $.

We shall first evaluate the entropy distribution as a function of
rapidity $y$ and time $t$ in the first stage of hydrodynamics.
Consider a slab element $dz$ at $z$ at a fixed time $t$.  The entropy
within the slab element is
\begin{eqnarray}
dS=\sigma u^0 dz.
\end{eqnarray}
Using the solution (\ref{sol2}), we can express $z$ as a function of
$t$ and rapidity $y$ during the one-dimensional longitudinal
expansion,
\begin{eqnarray}
\label{zt}
z=t\, {\sinh y}/{\cosh y}.
\end{eqnarray}
For a fixed value of $t$, we therefore obtain
\begin{eqnarray}
dS =  {\sigma  \, t \, dy}/{\cosh y} .
\end{eqnarray}
The entropy density $\sigma$ is related to $\epsilon$ by $\sigma =c
\epsilon^{3/4}$ and $\epsilon$ is given by (\ref{sole}).  We obtain
the rapidity distribution at the time $t$,
\begin{eqnarray}
dS &=& c \epsilon_0^{3/4} \exp\{ -(y_++y_- 
-\sqrt{y_+y_-} )\} \, {t ~dy}/{\cosh y}. 
\end{eqnarray}
In the second stage, different fluid elements with different
rapidities switch to conic flight at different time $t_{\rm FO}(y)$.  The
rapidity is frozen after $t> t_{\rm FO}(y) $.  The final rapidity
distribution after freeze-out needs to be evaluated at the switching
time $t=t_{\rm FO}(y) $
\begin{eqnarray}
\label{rap}
dS &=& c \epsilon_0^{3/4} \left [\exp\{ -(y_++y_- 
-\sqrt{y_+y_-}) \} \frac{t}{\cosh y} \right ]_{t=t_{\rm FO}(y) } dy
\end{eqnarray}
To evaluate the square-bracketed quantity at $t=t_{\rm FO}(y) $, we obtain from
Eq.\ (\ref{ypm}) and (\ref{zt}) that
\begin{eqnarray}
e^{y_\pm}=\frac{t}{\Delta} \frac{e^{\pm y}}{\cosh y}.
\end{eqnarray}
Therefore, we have
\begin{eqnarray}
\label{extm}
e^{y_\pm} \bigl | _{t=t_{\rm FO}(y) } =\frac{t_{\rm FO}(y) }{\Delta} \frac{e^{\pm y}}{\cosh y}
=\frac{2a}{\Delta}e^{\pm y} ,
\end{eqnarray}
which gives 
\begin{eqnarray}
\label{yyy}
y_{\pm} \bigl | _{t=t_{\rm FO}(y) }
= \ln \left ( {2a}/{\Delta}\right ) \pm y.
\end{eqnarray}
We note that 
\begin{eqnarray}
\label{ybb}
\ln \left ( {2a}/{\Delta} \right )=y_b=L+\ln 2,
\end{eqnarray}
where $y_b$ is the beam rapidity in the center-of-mass system,
\begin{eqnarray}
y_b = \cosh^{-1} \left ( {\sqrt{s_{NN}}}/{2m_p} \right )
\doteq \ln \left ( { \sqrt{s_{NN}}}/{m_p} \right ).
\end{eqnarray}
The rapidity distribution of Eq.\ (\ref{rap}) is therefore
\begin{eqnarray}
dS &=& c 
\epsilon_0^{3/4} 2 a \exp\{ -2y_b 
+\sqrt{y_b^2-y^2} \}dy
\end{eqnarray}
As the entropy is proportional to the number of particles, we obtain
the rapidity distribution
\begin{eqnarray}
\label{new}
{dN}/{dy} &\propto& \exp\{\sqrt{y_b^2-y^2} \}.
\end{eqnarray}
which differs from Landau's rapidity distribution of Eq.\ (\ref{Lan}).

While many steps of the formulation are the same, the main difference
between our formulation and Landau's appears to be the additional
factor of 2 in Eq.\ (\ref{extm}) and (\ref{tmy}) in the new
formulation.  This factor can be traced back to the factor of 4 in the
ratio of $4\epsilon/3$ from $(\epsilon+p)$ on the left hand side of
Eq. (\ref{displ}) and $\epsilon/3$ from the pressure $p$ on the right
hand side.  In Landau's formulation, this factor of 4 is taken to be
unity for an order-of-magnitude estimate of the transverse expansion.

\large
\vspace*{0.5cm} 
{\noindent 
\section {\bf Comparison of Landau
Hydrodynamics with Experimental Rapidity Distributions}
}
\vspace{0.3cm}

\begin{figure} [h]
\includegraphics[angle=0,scale=0.40]{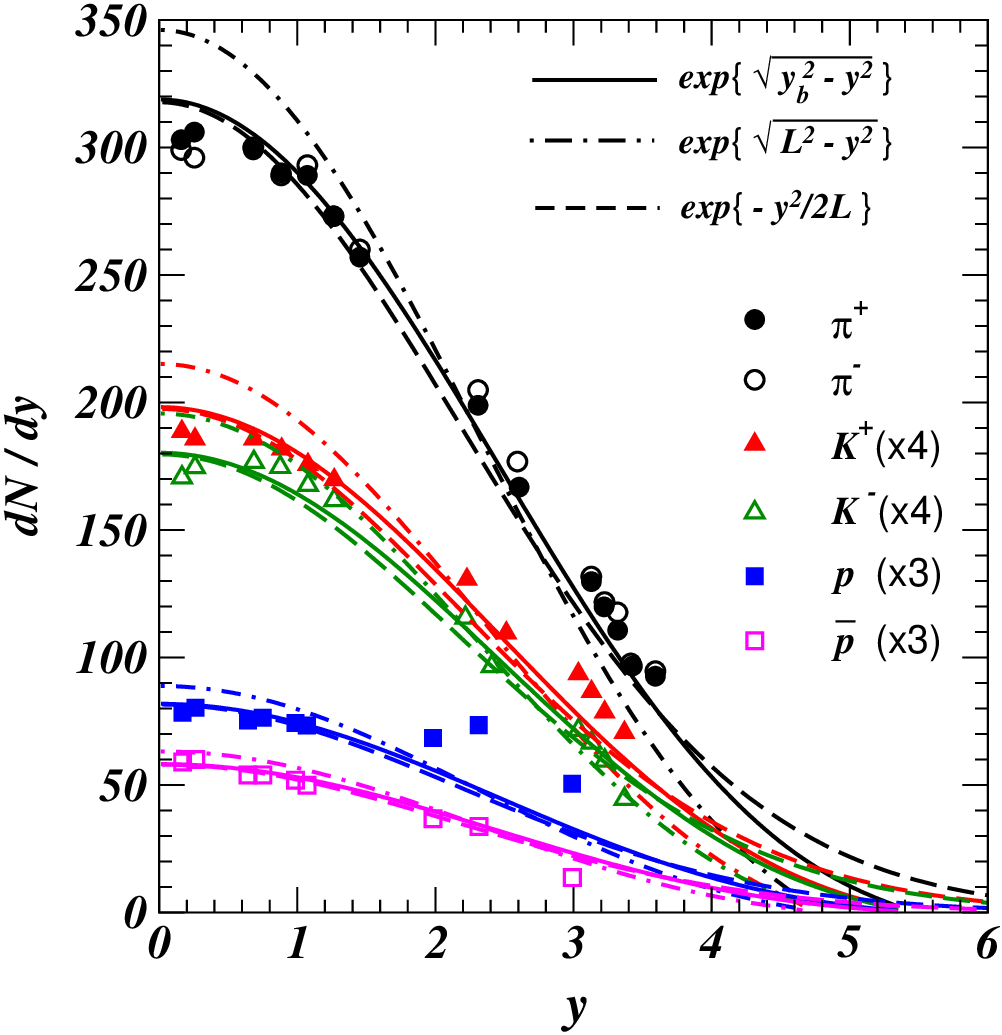}
\vspace*{0.0cm} 
\caption{(Color online)  Comparison of experimental rapidity distribution with
theoretical distribution in the form of $dN/dy \propto \exp\{
\sqrt{y_b^2-y^2} \}$ (solid curves), Landau's distribution $dN/dy({\rm
Landau}) \propto \exp\{ \sqrt{L^2-y^2} \}$ (dashed-dot curves), and
the Gaussian $dN/dy({\rm Gaussian}) \propto \exp\{-y^2/2L \}$ (dashed
curves) for produced particles with different masses. Data are from
\cite{Mur04} for AuAu collisions at $\sqrt{s_{_{NN}}}=200$ GeV.  }
\end{figure}

Fig.\ 3 gives the theoretical and experimental rapidity distributions
for $\pi^+$, $\pi^-$, $K^+$, $K^-$, $p$, and $\bar p$ at
$\sqrt{s_{_{NN}}}=200$ GeV \cite{Mur04}.  The beam rapidity is
$y_b=5.36$, and the logarithm of the Lorentz contraction factor is $
L=4.67$.  The solid curves give the the modified distribution
Eq.\ (\ref{new}), whereas the dashed curves are the Landau
distribution of Eq.\ (\ref{Lan}).  The theoretical distributions for
different types of particles have been obtained by keeping the
functional forms of the distribution and fitting an overall
normalization constant to the experimental data.  We observe that
Landau rapidity distributions are significantly narrower than the
experimental rapidity distributions, whereas the modified distribution
Eq.\ (\ref{new}) gives theoretical results that agree better with
experimental data.

As a further comparison, we show theoretical distributions calculated
with the Gaussian distribution of Eq.\ (\ref{gau}) as the dashed
curves in Fig.\ 3.  We find that except for the region of large
rapidities, the Gaussian distributions is a good representation of the
modified Landau distribution.  The close similarity between the
modified distribution (\ref{new}) and the Gaussian distribution
(\ref{gau}) explains the puzzle mentioned in the Introduction.  The
Gaussian distribution and the original Landau distribution are
different distributions.  Past successes of the Gaussian distribution
in explaining experimental rapidity data \cite{Mur04,Ste05,Ste07}
arise, not because it is an approximation of the original Landau
distribution (\ref{Lan}), but because it is in fact close to the
modified Landau distribution (\ref{new}) that derives its support from
a careful re-examination of Landau hydrodynamics.

\begin{figure} [h]
\includegraphics[angle=0,scale=0.40]{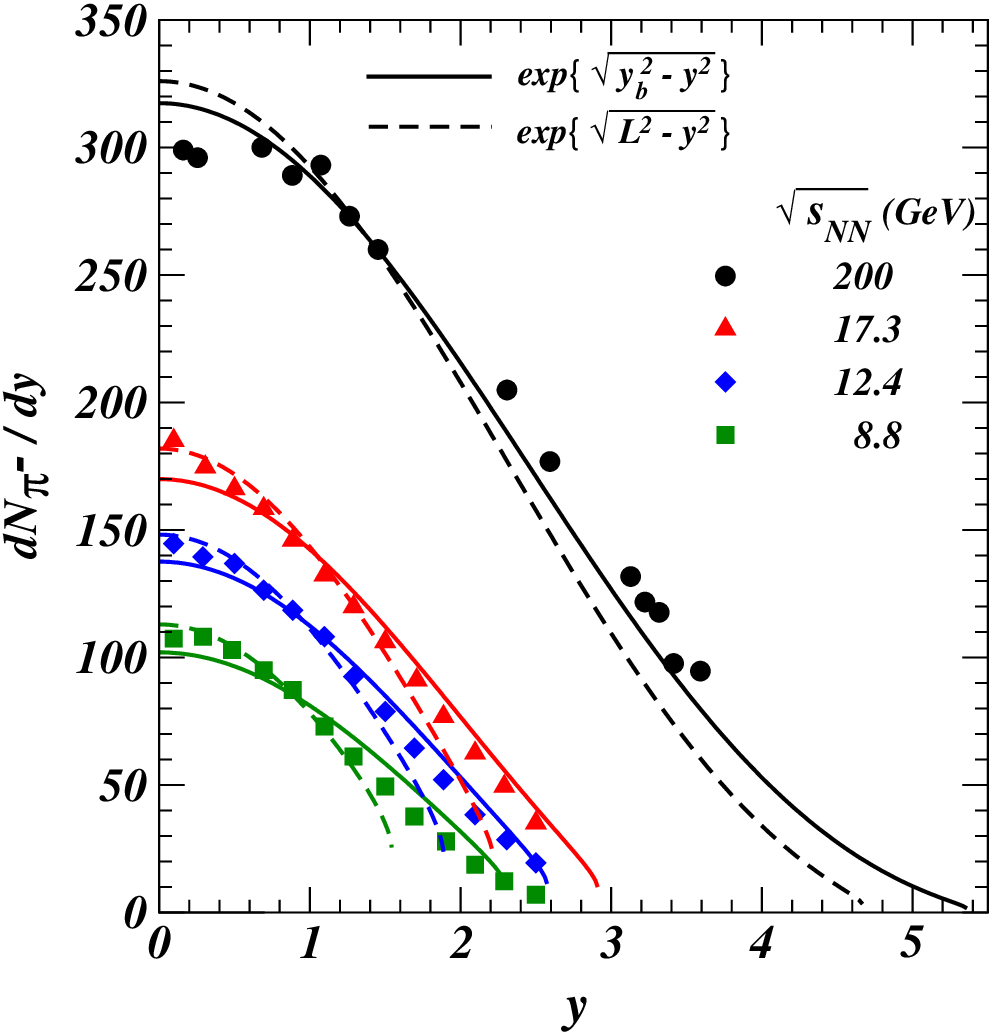}
\vspace*{0.0cm} 
\caption{(Color online) Comparison of experimental rapidity
  distribution with theoretical distribution in the form of $dN/dy
  \propto \exp\{ \sqrt{y_b^2-y^2} \}$ (solid curves) and Landau's
  distribution $dN/dy ({\rm Landau}) \propto \exp\{ \sqrt{L^2-y^2} \}$
  (dashed curves) for produced particles at different
  energies. Experimental $dN_{\pi^-}/dy$ data are from the
  compilations in \cite{Mur04}.  }
\end{figure}

We compare theoretical distributions with the $\pi^-$ rapidity
distribution for collisions at various energies.  The solid curves in
Fig.\ 4 are the results from the modified distribution Eq.\
(\ref{new}) with the $y_b$ parameter, whereas the dashed curves are
the Landau distribution of Eq.\ (\ref{Lan}) with the $L$
parameter. The experimental data are from the compilation of
\cite{Mur04}.  The modified distributions of Eq.\ (\ref{new}) give a
better agreement with experimental data than the original Landau 
distributions.

\vspace*{0.5cm}
\section
{\bf Predictions of Rapidity Distributions for LHC Energies}
\vspace{0.3cm}

We can re-write the rapidity distribution of charged
particles in terms of the normalized distribution $dF/dy$
\begin{eqnarray}
( dN_{\rm ch} / dy) / (N_{\rm part}/2) = 
 [N_{\rm ch} / (N_{\rm part}/2)] dF/dy. 
\end{eqnarray}
The normalized distribution $dF/dy$ is
\begin{eqnarray}
\label{shape}
\frac{dF}{dy} = \begin{cases} A_{\rm norm} \exp\{\sqrt{y_b^2-y^2}\} &
{\rm for~ modified~ distribution}, \\ A_{\rm norm}
\exp\{\sqrt{L^2-y^2}\} & {\rm for~ Landau~ distribution},\\
\frac{1}{\sqrt{2\pi L}} \exp\{{-y^2/2L}\} & {\rm for~
Gaussian~distribution},
\end{cases}
\end{eqnarray}
where $A_{\rm norm}$ is a normalization constant such that
\begin{eqnarray}
\int  dF/dy =1.
\end{eqnarray}

With the knowledge of the total charged multiplicity from Fig.\ 1, and
the shape of the rapidity distribution from Eq.\ (\ref{shape}), we can
calculate $dN_{\rm ch}/dy / (N_{\rm part}/2)$ as a function of
rapidity.  Fig. 5 gives the predicted rapidity distributions at LHC
energies.  For heavy-ion collisions at $\sqrt{ s_{_{NN}}}=$ 5.5 TeV
with full stopping, the maximum value of $dN/dy$ per participant pair
is about 22 at midrapidity.  For $pp$ collisions at $\sqrt{
s_{_{NN}}}=$ 14 TeV with $\xi=0.5$, the maximum $dN/dy$ is
approximately 24 at $y=0$.  The widths of the rapidity distributions
are $\sigma_y\sim 3$.  The solid curves are for the modified
distribution, the dashed-dot curves are for the original Landau
distribution, and the dashed curves are for the Gaussian distribution.
\begin{figure} [h]
\includegraphics[angle=0,scale=0.50]{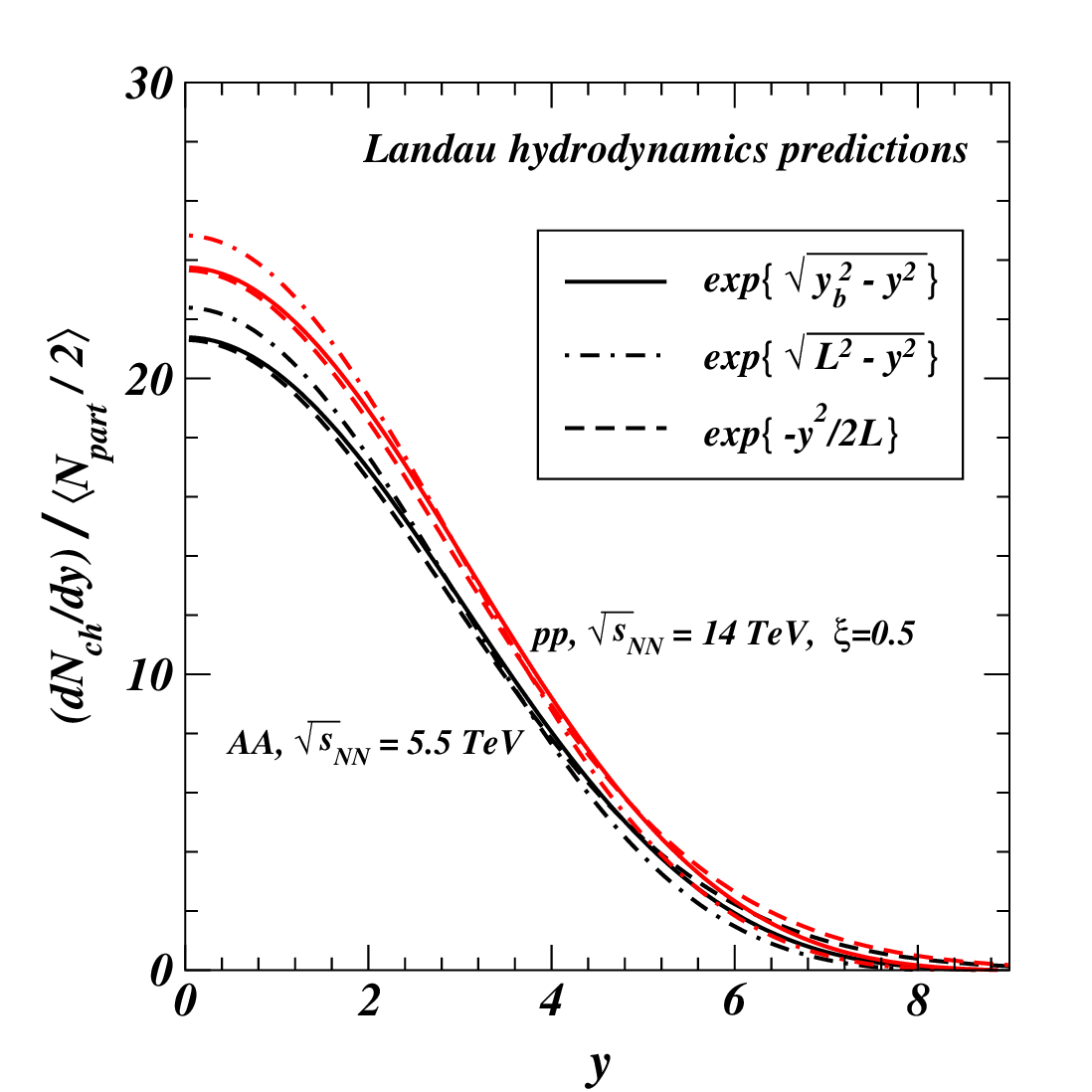}
\vspace*{0.0cm} 
\caption{(Color online) The predicted rapidity distributions $dN_{\rm ch}/dy /
 (N_{\rm part}/2)$ of charged particles produced in $pp$ collisions at
 $\sqrt{s_{_{NN}}}=14$ TeV with $\xi=0.5$, and AA collisions at
 $\sqrt{s_{_{NN}}}= 5.5$ TeV with full stopping in Landau
 hydrodynamics.  The solid curves are obtained with the modified
 distribution, the dashed-dot curves are obtained with the original
 Landau distribution, and the dashed curves with the Gaussian
 distribution.  }
\end{figure}

\section
{\bf Generalization to Non-Central Collisions}
\vspace{0.3cm}

In non-central collisions, the transverse radius $a_\phi/2$ will
depend on the azimuthal angle $\phi$ measured relative to the $x$ axis
as depicted in Fig. 2.  Following the same Landau arguments as in the
central collision case, Eq.\ (\ref{displ}) for the transverse
displacement can be generalized to be 
\begin{eqnarray}
\label{displrho}
\frac{4}{3} \epsilon u^0 u^0 \frac{ 2 \rho(\phi,t) }{ t^2}
= \frac{ \epsilon}{ 3a_\phi/2}. 
\end{eqnarray}
where $\rho(\phi,t)$ is the transverse displacement at azimuthal angle
$\phi$.  The transverse displacement depend on
$\phi$ and $t$ as
\begin{eqnarray}
 \rho(\phi, t) = \frac{t^2}{4 a_\phi u^0 u^0 }=\frac{t^2}{4 a_\phi
 \cosh^2 y} .
\end{eqnarray}
The Landau condition for the onset of the second stage
as the condition that the transverse displacement $\rho(\phi, t)$ is
equal to the transverse dimension $a_\phi$,
\begin{eqnarray}
 \rho(\phi, t_{\rm FO}) = a_\phi.
\end{eqnarray}
Thus, in the case of non-central collision, the Landau condition of
(\ref{xtt}) is changed to 
\begin{eqnarray}
t_{\rm FO}(y,\phi) = (a_\phi/a)\times 2 a \cosh y.
\end{eqnarray}
Following the same
argument as before, Eq.\ (\ref{yyy}) for the non-central collision
case becomes
\begin{eqnarray}
y_{\pm} \bigl | _{t=t_{\rm FO}(y,\phi) }
= \ln (a_\phi/a)+ \ln(2a/\Delta_b)  \pm y,
\end{eqnarray}
where the longitudinal thickness of the
initial slab $\Delta_b$ depends on the impact parameter $b$.
As a consequence, the rapidity distribution for this non-central
collision is
\begin{eqnarray}
\frac{dN}{dy} &\propto& \exp\{\sqrt{\ln(2a/\Delta_b) + \ln (a_\phi/a)]^2-y^2} \}.
\end{eqnarray}

\large
\vspace*{0.5cm}
{\noindent 

\section{Corrections to the Landau Model}

Results in the last few sections deal with the Landau model in its
traditional form. It is gratifying that gross features of many
measured quantities are reproduced well.  The Landau model with 
the modified distribution (\ref{new}) can be considered a good first
approximation.  Corrections and refinements are expected to be small
and need to be included as physical considerations and experimental
data demand.  In this respect, it is useful to examine two important
corrections arising from uncertainties in the initial configuration
and the final freeze-out condition.

The Landau model assumes that the initial configuration corresponds to
a disk of thickness $\Delta$=$({\rm nuclear~diameter}~a)/\gamma$ as
given by Eq.\ (3.1).  Landau's hydrodynamical expansion commences at
the end of the initial compression, with the formation of shock waves
already at hand.  However, the thickness of the initial compressed
shock waves arises from balancing energy and momentum following the
Rankine-Hugoniot boundary conditions across the shock front
\cite{Won74,Cse87}. The longitudinal thickness of the compressed
region (shock region) depends not only on the diameter $a$ of the
nuclei, but also on the equation of state and the collision energies.
Thus, although the initial nuclear diameter $a$ is an important
scaling parameter as used by Landau, the longitudinal thickness of the
compressed region may deviate from the Landau's estimate of $a/\gamma$
due to the equation of state and collision energy considerations.  The
equation of state at AGS energies is more dominated by baryons while
the equation of state form RHIC collisions will be dominated by gluons
and quarks.  How the effects of the speed of sound can affect the
rapidity distribution in the Landau model have been examined recently
by Bialas and his collaborators and by Mohanty and his collaborator
\cite{Moh03}. There is furthermore the possibility of a much more
extended longitudinal configuration in the initial stages of highly
relativistic collisions in the string rope description of the initial
longitudinal compression \cite{Mag01}.  In that description, the
extension will depend on the string tension of the rope between the
separating partons, as investigated by Magas and his collaborators
\cite{Mag01}.  The observed strong azimuthal anisotropy as represented
by the azimuthal Fourier $b_n$ coefficients of \cite{Won79} (or the
$v_n$ coefficients in the later notation of \cite{Vol96} for elliptic
flow \cite{Dan85,Oll92}) may indicate this extended initial state of
\cite{Mag01} and an initial longitudinal dimension greater than
Landau's estimate.

There is another important correction to Landau's initial longitudinal
thickness because of the spherical geometry of the nuclei.  Landau
model assumes a initial longitudinal thickness of $a/\gamma$ with a
nearly-uniform longitudinal distribution for a nucleus with a diameter
of $a$ in its own rest frame.  However, the longitudinal distribution
of a spherical nucleus is far from being uniform.  A longitudinally
uniform cylinder of the same volume in a transverse disk of diameter
$a$ will have a longitudinal thickness equal to $2a/3$, which is
substantially smaller than the value of $a$ assumed by Landau.  The
density distribution of a spherical nucleus is also not uniform in the
transverse direction, when it is projected transversely.

All these corrections due to shock wave compression and spherical
geometry are expected to scale with the nuclear diameter $a$.  We can
introduce phenomenologically a correction factor $C_{\rm init}$ to
represent the effects of these scaled corrections so that the
longitudinal thickness changes from $\Delta=a/\gamma$ to $\Delta'$
\begin{eqnarray}
\Delta \to \Delta' = C_{\rm init} \times a/\gamma.
\end{eqnarray}
Upon replacing $\Delta$ by $\Delta'$, we get from Eqs.\ (\ref{ybb}) and
(\ref{new}) that $dN/dy$ is modified to become
\begin{eqnarray}
\label{dndy1}
\frac{dN}{dy} \propto
\exp\{\sqrt{(y_b-\ln{C_{\rm init}})^2-y^2} \}.
\end{eqnarray}
Thus the thickness correction factor $C_{\rm init}$ leads to a
logarithmic correction to the parameter $y_b$ in Landau's distribution
(\ref{new}).  For example, the geometrical correction of $C_{\rm
init}$(geometrical)$\sim 2/3$ contribute to a positive value of
$(-\ln{C_{\rm init}})\sim 0.405$, and a more extended initial shock
wave region as in \cite{Mag01} will lead to a negative contribution to
$(-\ln C_{\rm init})$ and a narrower rapidity width.  There is thus an
interplay between the static geometrical effects and the dynamical
effects due to compression and string rope extension.

There is an additional complication arising to the approximate
freeze-out condition.  Landau's freeze-out condition of $\tau_{\rm
  FO}=2a$ comes from his argument on the magnitude of the transverse
displacement.  Landau's freeze-out surface is a space-like surface
with a normal pointing in the time-like direction.  Important
contributions on the freeze-out condition comes from Cooper and Frye
who used a fixed temperature freeze-out condition.  They found that
the freeze-out surface in this case contains both the space-like
portion and time-like portion \cite{Coo74}.  Another important
contribution comes from Csernai who used the Rankine-Hugoniot
conditions to describe the freeze-out boundary \cite{Cse87}.  In this
case the balance of the transport across the freeze-out surface lead
to a modification of the transport equation for freeze-out
\cite{Mag05}.  In the unified description of Hwa-Bjorken and Landau
hydrodynamics, Bialas and his collaborators examined various
freeze-out conditions for fixed $t$, $\tau$, and temperature $T$, and
compare them with the original Hwa-Bjorken and Landau results
\cite{Bia07}.  Using a new family of simple analytical hydrodynamical
solutions, Cs\"org\"o and his collaborators use the fixed temperature
condition for the freeze-out \cite{Cso08}.  The effects of the speeds
of sound and the freeze-out temperature on the rapidity distribution
in the Landau model have been investigated recently by Beuf and his
collaborators \cite{Beu08}.

While there are many possible freeze-out conditions, the successes of
Landau hydrodynamics suggests that Landau's freeze-out condition can
be a crude first approximation and the correction is likely to be
small and scale with the Landau freeze-out proper time $\tau_{\rm
  FO}\sim 2a$.  Phenomenologically it is therefore useful to introduce
a corrective freeze-out factor $C_{\rm FO}$ to replacing $\tau_{\rm
  FO}$ by $\tau_{\rm FO}'$,
\begin{eqnarray} \tau_{\rm FO} \to \tau_{\rm FO}'=C_{\rm FO} \times 2a.
\end{eqnarray}
From Eq.\ (\ref{ybb}), this modification of the freeze-out proper time
lead to a modification of the rapidity distribution from $dN/dy$ of
Eq. (\ref{new}) which becomes
\begin{eqnarray}
\label{dndy2}
\frac{dN}{dy} \propto
\exp\{\sqrt{(y_b+\ln C_{\rm FO})^2-y^2} \}.
\end{eqnarray}
Again, the correction factor $C_{\rm FO})$ leads to a logarithmic
correction to $y_b$.
\begin{figure} [h]
\includegraphics[angle=0,scale=0.50]{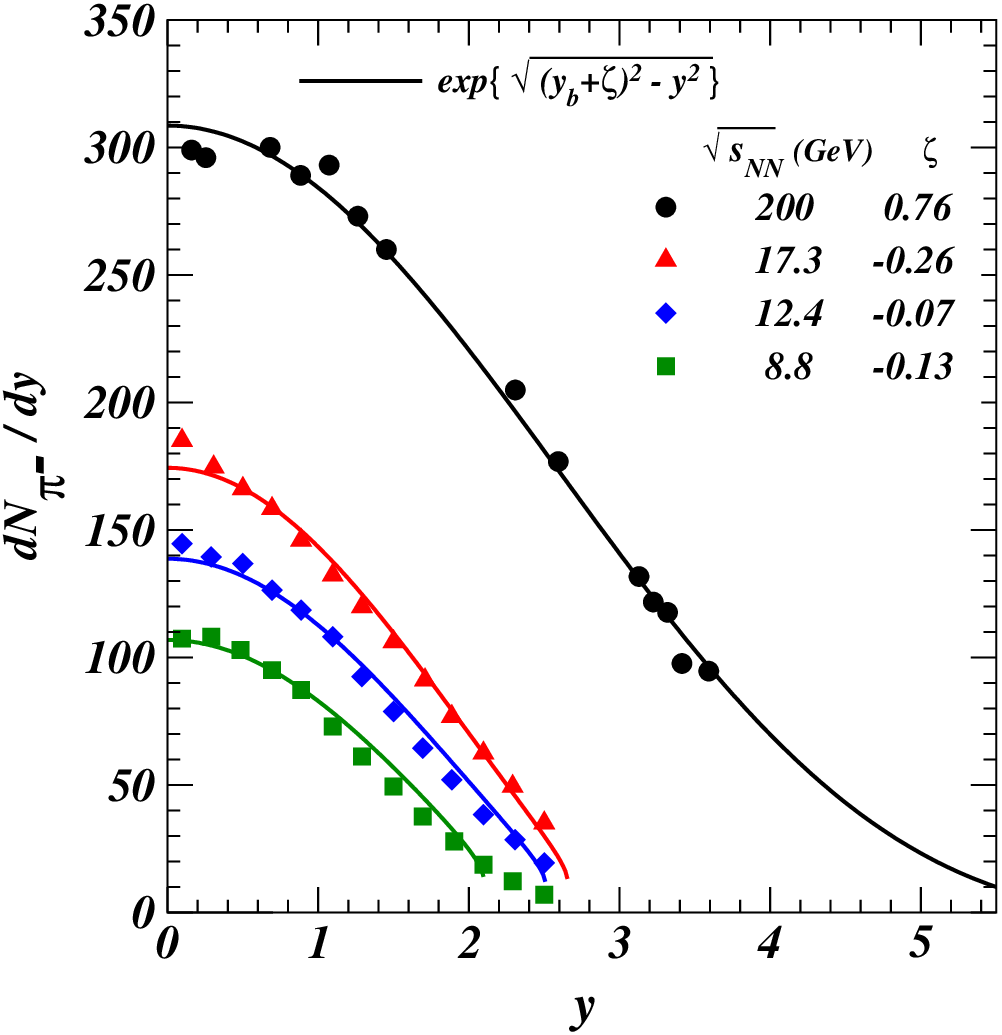}
\vspace*{0.0cm} 
\caption{(Color online) Comparison of experimental rapidity
distributions with theoretical rapidity distributions $dN_{ \pi^-}/dy$
(solid curves), calculated with Eq.\ (\ref{dndyzeta}) at various
energies.  Experimental data points are from \cite{Mur04}.  }
\end{figure}

The measured rapidity distribution depends on the combination of both
effects.  Upon combining the initial condition  and the
freeze-out condition 
corrections
from Eqs.\ (\ref{dndy1}) and  (\ref{dndy2}), we obtain
\begin{eqnarray}
\label{dndyzeta}
\frac{dN}{dy} \propto \exp\{\sqrt{(y_b+\zeta)^2-y^2} \}
\end{eqnarray}
where the correction parameter $\zeta$ is
\begin{eqnarray}
\zeta=-\ln C_{\rm init}+\ln C_{\rm FO}.
\end{eqnarray}
Our theoretical knowledge has not advanced to such an extent that we
can separate out the different effects due to the initial conditions
and the effects due to the freeze-out conditions as they closely
interplay to give rise to the observed rapidity distribution.  What is
possible is to extract the deviations of the experimental data from
the Landau model so that the small deviations may reveal useful
information in future investigation.  The agreement with experimental
$dN/dy$ data with theoretical predictions will be slightly improved
when we include this correction parameter $\zeta$.  In Fig.\ 6, we use
the experimental $dN_{\pi^-} /dy$ data \cite{Mur04} and the
distribution of Eq.\ (\ref{dndyzeta}) to extract the quantity $\zeta$
as a function of $\sqrt{s}$ shown in Table I.  As a comparison, the
corresponding values of $y_b$ are also listed.  One finds that for AGS
and SPS energies, the combined effects of initial and freeze-out
corrections lead to a small correction parameter $\zeta$ ranging from
-0.07 to -0.26.  The correction $\zeta$ is larger for RHIC energies
and assumes the value of 0.76.  In all cases, the magnitude of the
correction parameter, $|\zeta|$, is much smaller than $y_b$,
indicating the validity of the Landau model as a good first
approximation.  How the small correction $\zeta$ varies with collision
energy is an interesting topic worthy of future investigations.

The particle multiplicity in the Landau model is also affected by the
initial compressed volume and the bag constant as the observed
particles are hadrons subject to the bag pressure of confined quarks
and gluons \cite{Coo75}.  The effect of the bag constant is however
small for high-energy collisions \cite{Coo75}.

\vskip 0.5cm 
\centerline{Table I.  The correction parameter $\zeta$ as
a function of collision energy $\sqrt{s_{NN}}$.}  
\vskip 0.3cm
\centerline
{
\begin{tabular}{|c|c|c|}
\hline
${ \sqrt{s_{_{NN}}}}$  (GeV) &~~~~~~ $\zeta$ ~~~~~~&~~~~~~~~~$y_b$~~~~~~~~~\\
\hline
200  & ~0.76  &   5.362 \\
17.3 & -0.26  &   2.917 \\
12.4 & -0.07  &   2.575 \\
8.8  & -0.13  &   2.226 \\
\hline
\end{tabular}
}
\vskip 0.6cm

\section {\bf 
Conclusions and Discussions}}
\vspace{0.3cm}

In many problems in high-energy collisions such as in the description
of the interaction of the jet or quarkonium with the produced dense
matter, it is desirable to have a realistic but simple description of
the evolution of the produced medium.  Landau hydrodynamics furnishes
such a tool for this purpose.

Recent successes of Landau hydrodynamics in explaining the rapidity
distribution, total charged multiplicities, and limiting fragmentation
\cite{Mur04,Ste05,Ste07} indicate that it contains promising degrees
of freedom.  Questions are however raised concerning the use of
pseudorapidity or rapidity variables, the Gaussian form or the
square-root exponential form of the rapidity distribution, and the
values of the parameters in the rapidity distribution.

We start with the rapidity variable from the outset so that we do not
need to worry about the question of the rapidity or the pseudorapidity
variable.  We follow the formulation of the Landau hydrodynamics by
keeping careful track of the numerical constants that enter into the
derivation.  We confirm Landau's central results except that the
approximate rapidity distribution obtained by Landau needs to be
modified, when all numerical factors are carefully tracked.  In
particular, the rapidity distribution in the center-of-mass system
should be more appropriately given as $dN/dy \propto \exp \{
\sqrt{y_b^2-y^2}\}$, where $y_b$ is the beam nucleon rapidity, instead
of the Landau original result of $dN/dy({\rm Landau}) \propto \exp \{
\sqrt{L^2-y^2}\}$.  The modified distribution leads to a better
description of the experimental data, thereby supports the approximate
validity of Landau hydrodynamics as a description of the evolution of
the produced bulk matter.  

The modified distribution differs only slightly from the Gaussian
distribution $dN/dy({\rm Gaussian}) \propto \exp \{ -y^2/2L\}$, that
has been used successfully and extensively in the literature
\cite{Mur04,Ste05,Ste07,Car73}.  This explains the puzzle we mention
in the Introduction.  Even though the Gaussian Landau distribution
(\ref{gau}) is conceived as an approximate representation of the
original Landau distribution (\ref{Lan}) for the region of small
rapidity with $|y| \ll L$, it differs from the original Landau
distribution in other rapidity regions.  The Gaussian distribution has
been successfully used to explain experimental rapidity distribution
data \cite{Mur04,Ste05,Ste07}, not because it is an approximation of
the original Landau distribution (\ref{Lan}), but because it is in
fact a good representation of the modified Landau distribution
(\ref{new}) that derives its support from a careful re-examination of
Landau hydrodynamics.  Thus, there is now a firmer theoretical support
for the Gaussian distribution (\ref{gau}) owing to its similarity to
the modified distribution of (\ref{new}).

The need to modify Landau's original distribution should not come as a
surprise, as the original Landau distribution was intended to be
qualitative.  Our desire to apply it quantitatively therefore lead to
a more stringent re-examination, with the result of the modification
as we suggest.  The quantitative successes of the modified
distribution in Landau hydrodynamics make it a useful tool for many
problems in high-energy heavy-ion collisions.

In spite of these successes, many problems will need to be examined to
make the Landau model an even better tool.  We have discussed the
important effects of the initial configuration and the final
freeze-out condition in Section X.  They lead to uncertainties that
give rise to logarithmic correction parameter $\zeta$ with a magnitude
much smaller than $y_b$.  How the small correction $\zeta$ varies with
the collision energy is a subject worthy of further investigations.
We can also outline a few others that will need our attention.  The
distribution so far deals with flow rapidity of the fluid elements,
and the thermal distribution of the particles inside the fluid element
has not been included.  The folding of the thermal distribution of the
particles will broaden the rapidity distribution and should be the
subject of future investigations. Another improvement is to work with
a curvilinear coordinate system in the transverse direction to obtain
the transverse displacement.  This will improve the description of the
matching time in the transverse direction.  One may wish to explore
other forms of the freeze-out condition instead of Landau's transverse
displacement condition to see how sensitive the results can depend on
the freeze-out condition.  Finally, as the approximate solution for
the one-dimensional is also available, it may also be of interest to
see how much improvement there can be in obtaining the matching time
estimates that enter into the rapidity freeze-out condition.

\vspace*{0.3cm} The author wishes to thank Prof. D. Blaschke for his
hospitality at the the Helmholtz International Summer School, July
12-26, 2008, Bogoliubov Laboratory of Theoretical Physics, Dubna,
Russia, where this work on Landau hydrodynamics was initiated as
lecture notes.  This research was supported in part by the Division of
Nuclear Physics, U.S. Department of Energy, under Contract No.
DE-AC05-00OR22725, managed by UT-Battelle, LLC.

\end{document}